## RESEARCH ARTICLE

# Modeling and Characterizing Service Interference in Dynamic Infrastructures

VÍCTOR MEDEL[1], UNAI ARRONATEGUI[2], OMER RANA[3], JOSÉ ÁNGEL BAÑARES[2], AND RAFAEL TOLOSANA-CALASANZ[2]
[1]Nervia Consultores SL., 50001 Zaragoza, Spain
[2]Aragon Institute of Engineering Research (I3A), University of Zaragoza, 50018 Zaragoza, Spain
[3]School of Computer Science and Informatics, Cardiff University, CF 24 3AA Cardiff, U.K.

Corresponding author: Unai Arronategui (unai@unizar.es)

This work was supported in part by the Aragonese Government and the European Regional Development Fund "Construyendo Europa desde Aragón" through Computer Science for Complex System Modeling (COSMOS) Research Group under Grant T35_17D, and in part by the Spanish Program "Programa estatal del Generación de Conocimiento y Fortalecimiento Científico y Tecnológico del Sistema de I+D+i" under Project PGC2018-099815-B-100. The work of Víctor Medel was supported by the Fellowship from the Spanish Ministry of Economy. The work of Omer Rana was supported by the Engineering and Physical Sciences Research Council (EPSRC) through Privacy-Aware Cloud Ecosystems (PACE) Project under Grant EP/R033293/1 and Grant EP/R033439/1.

**ABSTRACT** Performance interference can occur when various services are executed over the same physical infrastructure in a cloud system. This can lead to performance degradation compared to the execution of services in isolation. This work proposes a Confirmatory Factor Analysis (CFA)-based model to estimate performance interference across containers, caused by the use of CPU, memory and IO across a number of co-hosted applications. The approach provides resource characterization through human comprehensible indices expressed as time series, so the interference in the entire execution lifetime of a service can be analyzed. Our experiments, based on the combination of real services with different profiles executed in Docker containers, suggest that our model can accurately predict the overall execution time, for different service combinations. The approach can be used by a service designer to identify *phases*, during the execution life-cycle of a service, that are likely to lead to a greater degree of interference, and to ensure that only complementary services are hosted on the same physical machine. Interference-awareness of this kind will enable more intelligent resource management and scheduling for cloud systems, and may be used to dynamically modify scheduling decisions.

## I. INTRODUCTION

Cloud computing has enabled the dynamic allocation of computational resources to support service execution by using resource abstractions such as Virtual Machines (VMs) and containers to deploy services. Container abstraction and mechanisms (e.g. Docker and Kubernetes) provide a low-overhead approach by providing a computational unit for executing an application within an isolated environment, and by reducing the start-up and termination times [1] of containers during resource provisioning, as well as physical resource usage, compared to VMs.

The associate editor coordinating the review of this manuscript and approving it for publication was Fabrizio Messina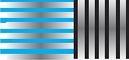.

Container and VMs have been viewed as complementary virtualization techniques that solve different problems [2]. While containers have been used as tools for the deployment of software with a Platform-as-a-Service (IaaS) focus, VMs have been considered the main hardware allocation and management tools with a focus on providing an IaaS. Initially, container technologies were adopted by enterprises to create homogeneous infrastructures over heterogeneous cloud computing solutions. In addition, the use of containers alleviates the problem of being locked into a cloud platform, due to lack of portability and interoperability between them [3], [4]. However, containers provisioned inside the VMs can exhibit performance degradation and unpredictable effects [5]. As companies seek better efficiency and lower cost in cloud deployments, they are increasingly interested in containers running on bare metal, instead of on VMs.

According to a recent IDC survey [6], 50 percent of the container workloads, in the enterprise market, are on premise infrastructure versus the public cloud. The same percentage share of workloads can be seen in local containers, on bare metal, as an alternative to virtual machines. This situation has fostered a new and growing market niche for running containers on bare metal servers (for instance, Equinix Metal and IBM Bare Metal Kubernetes) that is not sufficiently covered by large cloud service providers (although Amazon already offers some bare metal instances). The appearance of these new competitive cloud providers, and the interest of important cloud clients (Netflix), highlights the fact that most important commercial cloud platforms (Google, Amazon, or Alibaba) do not support built-in mechanisms to detect interference and mitigate its impact [7].

A comparative study by Sharma et al. [8] showed that containers running on bare metal exhibit better performance than VMs, but they can suffer from performance interference because they share the underlying OS kernel and, at least, some of the hardware resources (such as L3 cache). Multiple services executed on the same physical machine can lead to resource contention. When two or more services or tasks are executed over the same physical resource, *performance interference* can occur, leading to performance degradation compared to the execution of each service in isolation (i.e. on a dedicated resource). Various scheduling strategies [9], [10] attempt to avoid the co-existence of tasks that interfere with one another on the same physical resource. In Paragon [10] an interference-aware scheduler is proposed, which classifies a task based on how much interference it will cause to co-scheduled applications and how much interference it can tolerate for multiple shared resources. The main assumption of Paragon is that interference remains constant over the entire execution lifetime of a service.

In this study, we propose an innovative model of indices to estimate performance interference among containers, that host services running on bare metal, with the following features: (i) consideration is made for CPU, memory and I/O usage variation during execution (especially for long running services), and therefore an observation that performance interference is time dependent; (ii) recording of human-comprehensible indices for expressing resource usage during service execution. These indices are expressed as a time series that can be analyzed; (iii) Confirmatory Factor Analysis (CFA) [11] is used to identify how performance interference is affected by the aforementioned indices and service profiles. The modeling approach of multiple linear regression is subsequently used in combination with CFA to quantify the performance interference of co-scheduled services; (iv) the indices are general purpose, and can be used, with low overhead, by a scheduler to prioritize services that cause more/ less interference, and to limit interference on physical resources. The target platforms of our approach are bare metal providers, private clouds, public clouds and host machines, with support of processor performance counters. In addition, any long-running applications/services, sharing intense use of the same specific resources, are good candidates to improve their performance with our solution.

We validated our approach by conducting experiments using a set of applications: POV-Ray, IOzone, Stream, Metis, bzip2, pbzip2, Montage, blastn and blastx. These were chosen based on their different and relevant patterns of resource usage. Three of them, povray, iozone and stream, are used as primary benchmarking applications to estimate the interference between other applications. The experiments were executed using the Docker container framework, showing that the model can be successfully used to estimate performance interference and overall service execution time. The rest of this paper is organized as follows: Related work is discussed in Section II. In Section III, we define the concept of performance interference and analyze the factors that can exacerbate it. In Section IV, we describe the use of Confirmatory Factor Analysis (CFA) and how it can be applied in the context outlined in this study. Indices for characterizing our interference model are described in Section V, and experimental validation is carried out in Section VII. Based on the results in section VIII, we analyze how the methodology can be refined. Finally, the conclusions and future work are presented in Section IX.

## II. RELATED WORK

Container virtualization performance and a comparison with VM virtualization have been addressed in many studies, for instance [12]. These studies analyzed several performance metrics for different deployment configurations and workloads. In general, the proposed metrics are related to CPU, memory and storage performance and they describe how isolation between VMs and containers is different.

The problem of performance isolation, but not security isolation, is similar to interference. This problem has been studied in the context of cloud computing [13]. However, in [13], the focus was on multi-tenant applications without any consideration of the infrastructure on which such applications were deployed, and the results in this work were obtained by simulation. Prior studies have focused on the interference of VMs residing on the same physical host, particularly for cache memory and memory bandwidth resources [14], [15]. The majority of these studies use the Last Level Cache (LLC) counter to analyze memory behavior. The other resources analyzed are the I/O file system [16] and the communication network [17].

Performance interference, which arises when multiple VMs compete for shared physical machine, has a negative impact on the quality of service perceived by the end user. Commercial cloud platforms use intelligent schedulers, live migration, or server reconfigurations [18] to solve this problem. These techniques are useful for infrastructure providers but not for end consumer. Clients must develop their own mechanisms to ensure the Quality of Service [19].

The characterization of applications is usually performed, by assigning weights to denote the importance of the CPU or memory, in predicting interference [20], [21], [22], [23]. In the case of network-intensive services, interference can also be quantified by monitoring network utilization [24]. These measures can be used [7] to ensure that the response time target of a specified end user is met in this context.

In addition, performance degradation has been detected in VM virtualization owing to vCPU scheduling issues [25] that container virtualization cannot experience, because containers do not manage vCPUs. In [26], the performance isolation of VMs on the same physical host was analyzed and the authors provided performance estimations based on the similarity between applications. In [27], a technique called Bubble-Up was used to estimate the execution time under contingency conditions in VM clusters. In this study, each application is modeled as a sensitivity profile, which is normalized to a single score, called the bubble score. Additionally, Hubbub-Scale [28] calculates an interference index based on the value of the degradation. In ROSA [29], an unceRtainty-aware Online Scheduling Algorithm has been proposed, to schedule dynamic and multiple workflows with deadlines, to reduce the interference from uncertain task start/execution/finish time, as well as the uncertain data transfer time among tasks or the sudden arrival of new workflows.

Other studies have focused on understanding the interference with physical servers. In Paragon [10], a collaborative filtering algorithm was used to determine the influence of several ''Sources of Interference'' in applications running on exclusive servers on Amazon EC2 and servers on a private cluster. Using this information, the scheduler attempts to choose the optimal machine for an application. Unlike our work, this technique considers that container interference remains constant for the entire execution, which might lead to non-optimal decisions for the scheduling algorithm. In addition, this study assumes that the host machines in the cluster are heterogeneous, and no meaningful metric can be found, such as our interference profiles. The machine-learning models, explained in this work, provide several variables that do not have a direct correspondence with physical observed variables, which can make it difficult to analyze the model and use it to reason for different situations.

Furthermore, the values of our interference profiles, or the estimated degradation, may be used by the scheduler to give some priorities, or penalties, for certain hosts. For instance, in ARQ [30], the concept of the Quality of a Resource required by an application is introduced. Low values of this metric correspond to applications that are insensitive to interference. This metric can be used as a priority parameter for a scheduler.

It is important to note that interference in cloud environments can also affect security. Delimitrou and Kozyrakis [31] proposed a methodology to determine which applications can be co-scheduled based on the likely security interference between them.

In a previous work [1], we proposed a Petri Net performance model to analyze the overhead of deployment and termination of containers in Kubernetes, with different configurations of Kubernetes pods, each with different number of containers. However, the important and complex problem of resource contention, which arises when several long running services are executed in containers sharing the same physical node, had not been addressed. This present study proposes a new and original approach for this contention problem, in the context of a broader work on the use of models for resource management in cloud environments [32].

### III. CONTAINERS AND SOURCES OF INTERFERENCE

Containers decouple the execution of a service and its dependencies on the operating system and environment over which they are deployed. As several containers can be running on the same physical (or virtual) machine, this can degrade their performance, for example, when a resource allocated to one container is not released until the container has finished executing a task. However, the computational resources involved in the execution of a container on a physical node are complex. The interference between containers is measured as the performance loss caused by the execution of one container at the same time as another on the same host. This is the difference between the execution time when the container has all the available computational resources and the execution time when there are other containers using those resources, thereby leading to resource contention. In Section VI, we measure this metric as the ratio between the time an application takes to execute to a pre-determined set point when scheduled on its own, compared to when it is co-scheduled with another application. The following Sources of Interference (SoIs) can lead to resource contention between containers, each of which is related to the physical resource(s) hosting the container:

- **CPU usage**: In most container management systems, if there is no contention in the use of the CPU, each container uses the required CPU; otherwise, there is a scheduling mechanism to share the CPU proportionally.
- **Cache Memory** and **Memory bandwidth**: The cache hierarchy in a node is not isolated between containers; therefore, a container can continuously fail to access cache memory because another container is making aggressive use of the cache.
- **Network usage**: Access to the network is shared between all containers on a node, and if there is no contention, a container can use the entire available network bandwidth.
- **I/O file system** access: Like the network, the file system is shared between all containers; additionally, it could also be a distributed file system that use the network.

We focus on CPU and cache memory usage as they have the most significant impact on container performance, and because cache memory is very difficult to isolate in these environments. However, the analysis of the network and I/O behavior is straightforward following our proposed methodology. The main difference between these resources is that their effective analysis requires the measurement of other low-level events, including support for network synchronization between containers. CPU usage by a container can be easily isolated through a reservation mechanism. However, our studies show that this technique leads to an increase in the execution time of the container owing to the overhead introduced by the reservation mechanism [9], which is twice the total execution time. An ad-hoc solution is to overbook the required resources, but such considerations should be at design and not operational time.

## IV. RESEARCH METHOD

Confirmatory Factor Analysis (CFA) [11] is a set of statistical techniques that identifies how observed, or measured, variables are affected by a set of factors (or latent variables that cannot be measured). CFA is a subset of Structural Equation Modeling (SEM) techniques [33]. In SEM, more sophisticated relationships among variables are allowed, enabling the construction of hierarchical models to construct composite indices. We chose CFA because it allows a researcher to establish a hypothesis, and unlike other factor analysis techniques such as Exploratory Factor Analysis (EFA) [34], the statistical model confirms or rejects the hypothesized model. To extract meaningful factors, the solution should be rotated and an arbitrary cut-off value must be used to determine which variable influences each factor. On the other hand, Principal Component Analysis (PCA) [35] only reduces the observed variables into a set of fewer factors that are (often) difficult to interpret.

### A. CONFIRMATORY FACTOR ANALYSIS

Formally, given a set of $p$ observed variables, $X$, and a set of $m$ factors $F$, we expect Equation 1 to be satisfied. $\mu_i$ is the intercept for $x_i$, that is, it is the expected value when all factors are 0, $\epsilon_i$ is the stochastic error and $m < p$.

$$x_i = \mu_i + \lambda_{i1}F_1 + \lambda_{i2}F_2 + \ldots + \lambda_{im}F_m + \epsilon_i \quad (1)$$

If all observed variables are affected by a single factor, we call the model a measurement model. Namely,

$$(\forall i \in \{1, p\}, \exists k \in \{1, m\} \mid \lambda_{ik} \neq 0)$$
$$\wedge (\forall j \in \{1, m\}, j \neq k, \lambda_{i,j} = 0)$$

We can express Equation 1 in matrix form as in eq. 2.

$$X = \begin{pmatrix} x_1 \\ x_2 \\ \vdots \\ x_p \end{pmatrix} \Lambda = \begin{pmatrix} \lambda_{11} & 0 & \cdots & 0 \\ \lambda_{21} & 0 & \cdots & 0 \\ \vdots & \vdots & \ddots & \vdots \\ 0 & 0 & \cdots & \lambda_{p-1m} \\ 0 & 0 & \cdots & \lambda_{pm} \end{pmatrix}$$

$$F = \begin{pmatrix} F_1 \\ F_2 \\ \vdots \\ F_m \end{pmatrix} \mu = \begin{pmatrix} \mu_1 \\ \mu_2 \\ \vdots \\ \mu_p \end{pmatrix} \epsilon = \begin{pmatrix} \epsilon_1 \\ \epsilon_2 \\ \vdots \\ \epsilon_p \end{pmatrix}$$

$$X = \Lambda F + \mu + \epsilon \quad (2)$$

If we suppose that $\Sigma = \text{Cov}(X - \mu)$, and we denote the covariance over the factors by $\Phi$, and the covariance over the error by $\Psi$, we can write Equation 2 in covariance form (Equation 3).

$$\Sigma = \text{Cov}(X - \mu) = \text{Cov}(\Lambda F + \epsilon)$$
$$= \Lambda \text{Cov}(F)\Lambda^t + \text{Cov}(\epsilon)$$
$$= \Lambda \Phi \Lambda^t + \Psi \quad (3)$$

To identify the model, we set the scale of latent factors. Two methods can be used: (i) fix the loading of the first observed variable for each factor to 1; or (ii) fix the factor variance to 1, i.e. $\forall i \in \{1, \ldots, m\}, \sigma(F_i) = 1$ – there are different methods in literature to estimate parameters $\Lambda$ and $\Phi$ [36]: (a) Maximum Likelihood (ML), (b) robust ML (MLR) and, (c) Weighted Least Squares (WLS). These methods are available in a number of statistical frameworks, such as R, STATA and MPLUS. In this work, we used the Maximum Likelihood (lavaan package [37] in the R statistical software[1]).

### B. FACTOR SCORES

Several methods can be used to compute the value (factor score) of latent variables [38]. In this sutdy, we use regression-based Thurstone or Thompson scores. Once the $\Lambda$ and $\Phi$ values are estimated, we can compute the Factor scores using regression (Equation 4) – matrix $B$ has coefficient values obtained using regression.

$$F = BX \quad (4)$$

In the general formulation of the CFA problem (Equation 1), we can include $n$ sampled values for the $p$ observed variables (Equations 5 and 6).

$$\begin{pmatrix} x_{11} - \mu_1 & \cdots & x_{1n} - \mu_1 \\ x_{21} - \mu_2 & \cdots & x_{2n} - \mu_2 \\ \vdots & \cdots & \vdots \\ x_{p1} - \mu_p & \cdots & x_{pn} - \mu_p \end{pmatrix}$$
$$= \begin{pmatrix} \lambda_{11} & 0 & \cdots & 0 \\ \lambda_{21} & 0 & \cdots & 0 \\ \vdots & \vdots & \ddots & \vdots \\ 0 & 0 & \cdots & \lambda_{p-1m} \\ 0 & 0 & \cdots & \lambda_{pm} \end{pmatrix} \times \begin{pmatrix} f_{11} & \cdots & f_{1n} \\ f_{21} & \cdots & f_{2n} \\ \vdots & \cdots & \vdots \\ f_{m1} & \cdots & f_{mn} \end{pmatrix} \quad (5)$$

$$\hat{X}_{p \times n} = \Lambda_{p \times m} \hat{F}_{m \times n} \quad (6)$$

[1]https://www.r-project.org/

We can isolate $F$ in 6 to compute $B$ as follows:

$$\hat{X} = \Lambda \hat{F} \implies \hat{F} = (\Lambda^t \Lambda)^{-1} \Lambda^t \hat{X}$$

As we want to minimise the error $\epsilon$, from Equation 3, we get:

$$\hat{\Sigma} = \Lambda \Phi \Lambda^t \implies (\Lambda^t \Lambda)^{-1} \Lambda^t = \Phi \Lambda^t \hat{\Sigma}^{-1}$$

Combining the previous expressions, we obtain Equation 7, where $\Phi$ is the covariance matrix across all Factors, $\Sigma$ is the covariance matrix across all observed variables and $\hat{X}$ is the matrix with the observed values centred on 0. $f_{ij}$ represents the factor score of $F_j$ for the $i^{th}$ observation.

$$\begin{aligned} \hat{F} &= \Phi \Lambda^t \hat{\Sigma}^{-1} \hat{X} \\ B &= \Phi \Lambda^t \hat{\Sigma}^{-1} \end{aligned} \quad (7)$$

## V. DEVELOPING INTERFERENCE INDICES

We characterized the behavior of service execution, identifying how different computational resources (cache, CPU, RAM, etc.) are used, to obtain meaningful indices that can allow the prediction of interference. Our interference indices correspond to the factors in our CFA model. To develop these indices: i) we conducted several experiments to obtain a dataset that includes different performance metrics that vary over the execution lifetime of several relevant services (subsection V-A); ii) we extracted several variables, from the measures in this dataset, which are correlated (Subsection V-B); iii) we carried out a CFA analysis to summarize these variables into four uncorrelated factors, which represent our proposed interference indices (Subsection V-C). Figure 1 illustrates how an application is characterized using these indices, as explained in more detail in the following subsections.

### A. BUILDING THE DATASET

First, we need to obtain a set of relevant measurements of resource usage to establish the observed variables for the CFA model. For this purpose, we chose a set of applications, based on their different and relevant patterns of resource usage, whose execution can help to build dataset of the relevant measurements, from real scenarios:

- **POV-Ray** v3.7 [39] is a ray tracing application that generates an image from a scene description. The application is multi-threaded and has high CPU requirements. The input parameters were provided by the default benchmark.
- **IOzone** [40] is a file system benchmark utility, and is executed with the following input parameters *iozone -a -i 0 -i 1 -g 4M*. Parameters have been chosen to adjust the execution time of the generated job. It has high cache hierarchy requirements with many cache misses.
- **Stream** [41] is a benchmark for testing the memory bandwidth of an architecture using a large input data array. To adjust the execution time of the task, the following parameters were used: *-DSTREAM_ARRAY_SIZE=100000000 -DNTIMES=100*.
- **Metis** [42] is a set of graph partitioning tools. We use the `gpmetis` tool with one iteration and two partitions, utilising the LiveJournal dataset,[2] which contains 4847571 nodes and 68993773 vertices.
- **Bzip2** and **pbzip2** – Bzip2 is an open-source file compression program and pbzip2 is a parallel version of this utility. The experiment compresses the LiveJournal dataset [2], which is approximately 1GB.
- **Montage** is a widely used scientific workflow that integrates several astronomical images into a single image mosaic. This task has high computational and data requirements. We used the Montage toolkit[3] with the provided *pleiades* example. We consider the entire pleiades workflow as a single task.
- **Blastn** and **Blastx** benchmarks[4] are related to the bioinformatics domain, involving string comparison between a target string and a large database of variable sized string sequences. They consist of several queries (100 for blastn and 24 for blastx) in a database to find biological sequences that resemble the original query string.

Additionally, we will use three of them, pov-ray, iozone and stream, as benchmarking applications to estimate the interference between applications (see Section VI), because each of them has a high usage of a specific resource (CPU, cache hierarchy and memory bandwidth).

The applications were hosted using containers with Docker platform. We have not used any CPU reservation technique to avoid the financial cost associated with this technique and some performance degradation [9]. We executed these containers and measured these variables at different points in their execution life-cycle. The objective is to obtain a dataset that captures the variations of the metrics in time series to build meaningful indices. We assume that when the application is started, all required data or information is available. Therefore the application does not need to communicate with other applications for input, and is primarily dependent on available physical resources.

As described in Section III, in this work we focus on CPU and cache memory use by applications, as relevant examples of resources whose use could lead to interference. The dataset consists of a number of observations of 11 variables that measure low-level resource usage of these applications executed inside a container. The metrics were measured with *perf* tool [43], which is available in several Linux distributions. This tool uses Linux performance counters to measure several hardware events. Given an application $A$ and a sampling period of time $s_A$ in seconds, we measure the number of occurrences (denoted by the # notation) of the following events:

[2] https://snap.stanford.edu/data/soc-LiveJournal1.html
[3] http://montage.ipac.caltech.edu/
[4] http://fiehnlab.ucdavis.edu/staff/kind/collector/benchmark/blast-benchmark

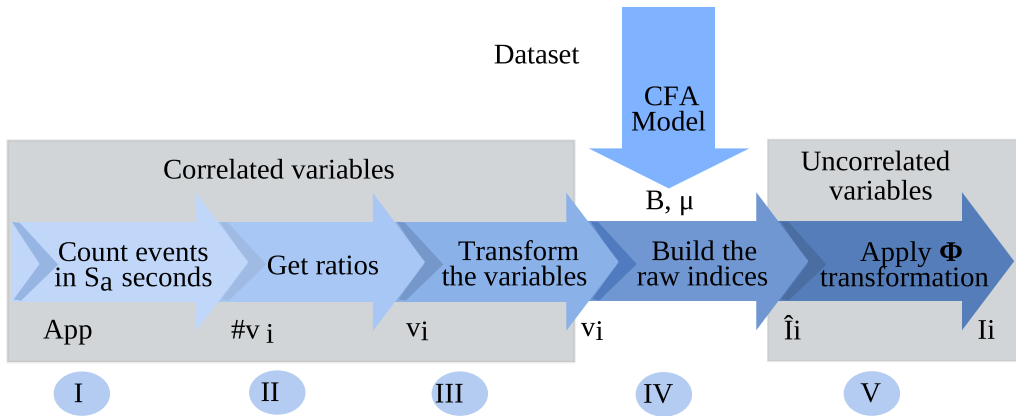

**FIGURE 1.** Process to get the interference indices from an application.

- **#Cycles**: Number of processor cycles executed in $s_A$ seconds. This is an indirect measure of CPU usage during this period.
- Number of **Cache-references** (#$C_r$) and **Cache-misses** (#$C_m$). They indicate the total cache accesses and misses from the memory hierarchy in $\hat{T}$ seconds. They reflect memory usage intensity and access patterns.
- **LLC-loads** (#$LLC_l$) and **LLC-load-misses** (#$LLC_{lm}$). They indicate the number of Last Level Cache accesses and misses for loading data in $s_A$.
- **LLC-stores** (#$LLC_s$) and **LLC-store-misses** (#$LLC_{sm}$). They indicate the number of Last Level Cache accesses and misses for storing data in $s_A$.
- **Branch-instructions** (#$B_r$) and **Branch-instructions-misses** (#$B_m$). They measure the number of instructions that cause the execution of a different instruction sequence and the number of instructions that cause a miss in the cache hierarchy.
- **Page faults** (#$fault$) measures the number of page faults when running the application. A page fault arise when the application attempts to access a virtual memory address that is not loaded into the physical memory. This event represents the lowest level of memory that requires access to the disk I/O.
- Number of executed **Instructions** (#$Inst$) counts the number of instructions executed in $s_A$ seconds.

### B. VARIABLES FROM THE DATASET

To obtain the measured variables to build the CFA model, instead of working with the absolute value of the number of occurrences of events presented in the previous subsection, we use their relative values as follows:

- Cache-references per instructions ($v_1$), branch-instructions per instructions ($v_2$), LLC-loads per instructions ($v_3$), and LLC-stores per instructions ($v_4$). These variables are obtained by dividing the number of events by the number of instructions executed over that period (Equation 8).

$$v_1 = \frac{\#C_r}{\#Inst} \qquad v_2 = \frac{\#B_r}{\#Inst}$$
$$v_3 = \frac{\#LLC_l}{\#Inst} \qquad v_4 = \frac{\#LLC_s}{\#Inst} \qquad (8)$$

- Cache-miss rate ($v_5$), branch-miss rate ($v_6$), LLC-load-miss rate ($v_7$) and LLC-store-miss rate ($v_8$). These variables are obtained by dividing the number of missed events by the number of their corresponding non-miss events executed over that period (Equation 9).

$$v_5 = \frac{\#C_m}{\#C_r} \qquad v_6 = \frac{\#B_m}{\#B_r}$$
$$v_7 = \frac{\#LLC_{lm}}{\#LLC_l} \qquad v_8 = \frac{\#LLC_{sm}}{\#LLC_s} \qquad (9)$$

- Standardized faults ($v_9$). We consider that this is not representative of the number of fault events per instructions executed. Instead, we standardize the value to avoid scaling problems in further analysis (Equation 10). The mean and standard deviation are calculated from the entire dataset, which captures the level of variability in the occurrence of these faults over time.

$$v_9 = \frac{\#faults - \mathrm{E}[\#faults]}{\mathrm{Var}[\#faults]} \qquad (10)$$

- CPU usage ($v_{10}$). This variable identifies the extent to which a CPU is used. Given the number of cycles executed in $s_A$ seconds, #Cycles, the number of cores of the machine, #Cores, and the CPU speed measured in MHz S, Equation 11 shows how the CPU usage is computed. The values lie within the [0, 1] interval.

$$v_{10} = \frac{\#Cycles}{s_A \cdot S \cdot 10^6 \cdot \#Cores} \qquad (11)$$

We used the $v_i$ notation to simplify the index expressions in the following sections.

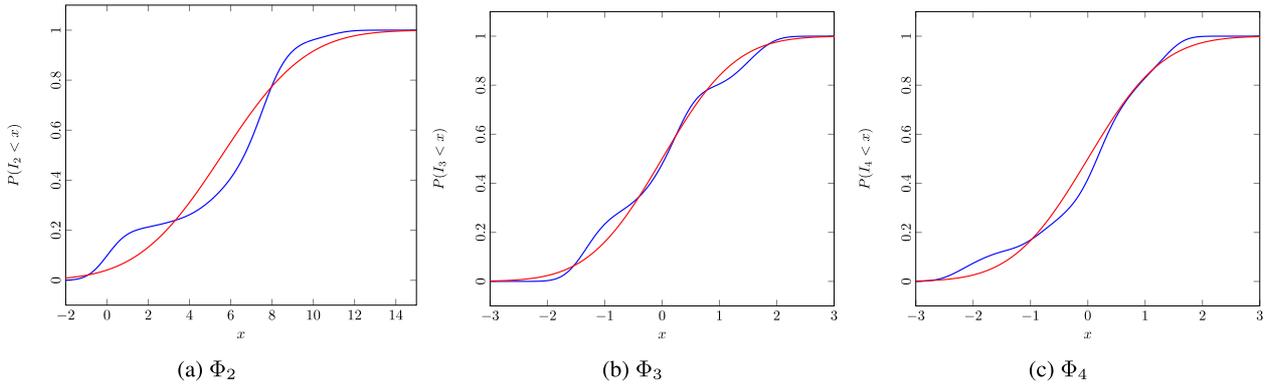

**FIGURE 2.** $\Phi_2$ –Subfigure (a)–, $\Phi_3$– Subfigure (b)– and $\Phi_4$ – Subfigure (c)– functions. Red line is the cumulative distribution function (cdf) of a normal distribution with the same mean and standard deviation.

### C. INTERFERENCE INDICES

The high correlation between the previous variables avoids their use as raw values to describe an application and perform further analysis. Therefore, to address meaningful and non correlated indices, we propose a theoretical model that combines these variables into the following four factors, expressed as indices:

- $I_1$. **CPU usage** index: This index models the intensity of CPU usage of an application.

$$I_1 = v_{10} \qquad (12)$$

- $I_2$. **Memory page fault**: This index models the number of page faults in the system. We analyzed them in isolation because they have significant impact on system performance. The observed values exhibited an exponential distribution. To improve the linearity of $I_2$, we transform the variable by taking natural logarithms (Equation 13).

$$I_2 = \Phi_2(\hat{I}_2), \text{ where}$$
$$\hat{I}_2 = \ln(v_9 + 1) \qquad (13)$$

- $I_3$. **Intensity of memory hierarchy usage** index: This index represents the number of accesses to memory hierarchy. It is a combination of the cache references –$\#C_r$, $\#LLC_l$– store references –$\#LLC_s$– and branch references –$\#B_r$– (Equation 14).

$$I_3 = \Phi_3(\hat{I}_3), \text{ where}$$
$$\hat{I}_3 = \sum_{i=1}^{i=4} b_i(\ln(v_i) - \mu_i) \qquad (14)$$

- $I_4$. **Intensity of cache misses** index: This index measures how frequently (in time and size) the application misses access to data in the cache hierarchy.

$$I_4 = \Phi_4(\hat{I}_4), \text{ where}$$
$$\hat{I}_4 = \sum_{i=5}^{i=8} b_i(\ln(v_i) - \mu_i) \qquad (15)$$

To avoid linearity problems in further analysis, we took logarithms for $\hat{I}_2$, $\hat{I}_3$ and $\hat{I}_4$ because they can be analyzed as ratios [44]. Because of the CFA formulation (Equation 1), we must center the variable before multiplying by its load; consequently, we subtract its mean value. It is important to note that $\hat{I}_2, \hat{I}_3, \hat{I}_4$ are values with a certain mean and standard deviation, and follow a non-normal probability distribution (theoretically, their values belong to the interval $(-\infty, \infty)$). We can transform these indices into other indices –$I_2, I_3, I_4$– through their cumulative distribution function (cdf). The transformation function $\Phi_X$ is defined as shown in Equation 16.

$$\Phi_X(x): \quad x \rightarrow P(X \le x)$$
$$(-\infty, \infty) \rightarrow [0, 1] \qquad (16)$$

We can build functions $\Phi_2$, $\Phi_3$, $\Phi_4$ empirically using their histogram, as shown in Figure 2. The raw indices do not follow a normal distribution, as denoted by the red line in Figure 2.

To compute the weight of each variable for $I_3$ and $I_4$, we performed CFA analysis as explained in Section IV. We fixed the variance of the latent factors to 1 –$\sigma(\hat{I}_2) = \sigma(\hat{I}_3) = 1$– and the covariance among factors to 0 – $\text{Cov}(\hat{I}_2, \hat{I}_3) = 0$–, thus, the resulting factors are forced to be orthogonal. The solution and further factor scores are calculated using the Lavaan package [37] on R. The parameter estimation method used is ML and the regression method is used to compute the scores.

The resulting $\mathbf{\Lambda}$, $\mathbf{\Phi}$ and $\mathbf{\mu}$ matrices are:

$$\mathbf{\Lambda} = \begin{pmatrix} \lambda_{11} & 0 \\ \lambda_{21} & 0 \\ \lambda_{31} & 0 \\ \lambda_{41} & 0 \\ 0 & \lambda_{52} \\ 0 & \lambda_{62} \\ 0 & \lambda_{72} \\ 0 & \lambda_{82} \end{pmatrix} = \begin{pmatrix} 2.202 & 0 \\ 0.179 & 0 \\ 2.342 & 0 \\ 2.011 & 0 \\ 0 & 1.315 \\ 0 & -0.259 \\ 0 & 1.241 \\ 0 & 1.483 \end{pmatrix}$$

$$\mathbf{\Phi} = \begin{pmatrix} \sigma(F_1) & \text{Cov}(F_2, F_1) \\ \text{Cov}(F_1, F_2) & \sigma(F_2) \end{pmatrix} = \begin{pmatrix} 1 & 0 \\ 0 & 1 \end{pmatrix}$$

$$\boldsymbol{\mu} = \begin{pmatrix} \mu_1 \\ \mu_2 \\ \mu_3 \\ \mu_4 \\ \mu_5 \\ \mu_6 \\ \mu_7 \\ \mu_8 \end{pmatrix} = \begin{pmatrix} -6.079 \\ -2.049 \\ -6.448 \\ -8.734 \\ -2.051 \\ -3.796 \\ -2.083 \\ -1.857 \end{pmatrix}$$

Then, we computed the $B$ matrix using Equation 19:

$$\mathbf{B} = \begin{pmatrix} 1 & -0.06 & -0.48 & -0.04 & 0 & 0 & 0 & 0 \\ 0 & 0 & 0 & 0 & 1.40 & 0.05 & -0.46 & -0.14 \end{pmatrix}$$

We can see that the values corresponding to $v_2$, $v_4$ and $v_6$ are quite low. Hence, they can be excluded from the analysis. The reason for this behavior is the high correlation between variables and because we are looking for factors that are not correlated. The variance in $v_2$ can be explained by other variables. Some of the observed variables (e.g. $v_3$ and $v_4$) measure events that might be included in others (e.g, $v_1$); thus, the scores of the factor remove the commonly measured elements and the correlation. Consequently these coefficients should not be interpreted as decreasing the interference caused by using a certain resource. In a computational environment, it is difficult to isolate these variables and to control them for a given application, and an analysis of the elasticity of the variables in the indices should not be carried out.

### D. CHARACTERIZING RESOURCE USAGE OF APPLICATIONS OVER TIME WITH INTERFERENCE INDICES

We want to model how the resource usage of an application changes over time using the four indices presented in the previous section, as time series. Given a sampling time $s_A$, to compute the value of $I_1$, $I_2$, $I_3$ and $I_4$ in that interval, we proceed as shown in Figure 1: i) we execute the containerized application alone to obtain the raw events; ii) we compute the ratios of the events; iii) we take logarithms in the $\hat{I}_2$ and $\hat{I}_3$ dependent variables; iv) we compute the factor scores for $\hat{I}_3$ and $\hat{I}_4$ through Equation 19; and v) we standardize and transform the variables to a [0, 1] range, using Equation 16. Steps 2 through 5 are repeated for each set of events measured with a period of $s_A$ seconds.

Figure 3 depicts the profile of pov-ray, iozone and stream applications. These applications are used in the following sections as benchmarks to analyze the impact of application co-scheduling, because they make a high usage of a certain resource. This situation is modeled by getting a high value on a certain index. For example, pov-ray –Figure 3a– continuously uses the entire computational capacity of the machine, therefore $I_1$ is about one for the entire execution. On the other hand, iozone – Fig 3b– makes random accesses to the memory, hence $I_4$ is about one. Stream application –Figure 3c– is a benchmark which tries to test the memory bandwidth, so the highest index is $I_3$.

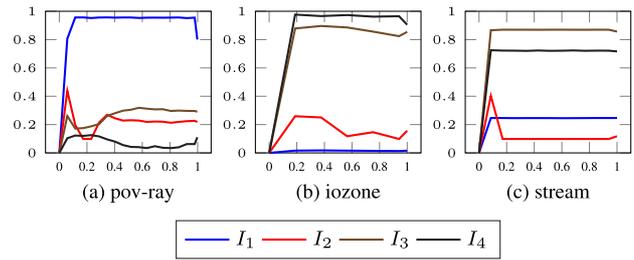

**FIGURE 3.** $I_1$, $I_2$, $I_3$ and $I_4$ values for benchmarking applications –(a) pov-ray, (b) iozone and (c) stream–. The execution time (x-axis) is normalized.

The behavior of the remaining applications is shown in Figure 4. We can see that pbzip2 –Figure 4d–, the parallel implementation of bzip2, makes a high usage of all resources in the system. When compared with the bzip2 profile (figure 4c) the usage of extra CPU capacity leads to an increase in faults in cache, $I_4$ index, while $I_3$ remains constant. Another interesting application is montage (figure 4b) which consists of several tasks executed in a pipeline scheme. This situation causes the memory usage indices, $I_3$ and $I_4$, feature some kind of pattern in their distribution. Similarly, blastn application (figure 4e) shows high values for $I_2$.

### VI. INTERFERENCE ANALYTICAL MODEL

In this section, we propose a model to estimate the interference between containers and, consequently, the total execution time under contingency situations. Interference is caused by sharing physical resources that are not isolated at the container level. To describe an application and its use of physical resources, we propose an index-based representation in Section V. We assume that when two applications are scheduled to be executed on the same machine, and if both use the same resource heavily, the degradation will be higher. Additionally, as the indices vary over the execution time, the interference will not be constant.

We propose a multiple linear regression model to estimate the interference between containers. Given an application $A$, we execute it with three benchmark applications that have a high usage of a particular resource to obtain reference interference values. These values allow us to build the regression model to estimate the interference when application $A$ is co-scheduled with another application whose resource profile is known.

The proposed methodology (Figure 5) comprises the following phases: (i) obtaining the interference profiles; (ii) benchmarking the application; (iii) defining the regression model; and (iv) estimating the interference. These phases are described in the following subsections.

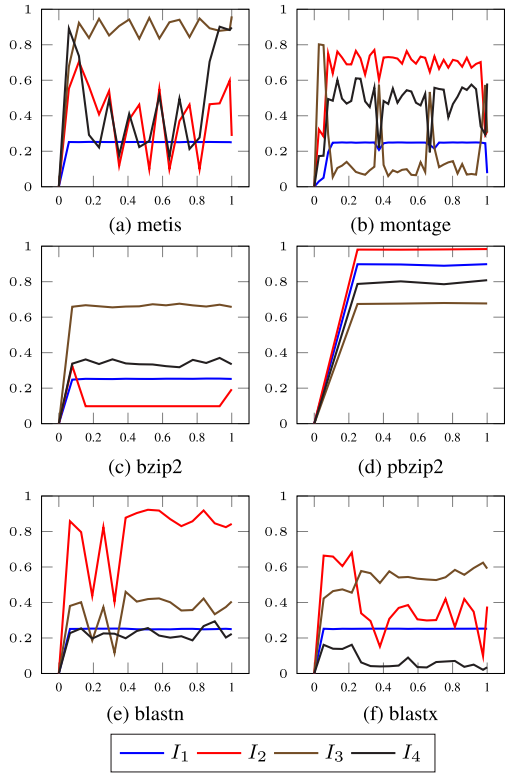

**FIGURE 4.** $I_1$, $I_2$, $I_3$ and $I_4$ values for some applications –(a) metis, (b) montage, (c) bzip2, (d) bzip2, (e) blastn and (f) blastx. The execution time (x-axis) is normalized.

### A. PRELIMINARIES

We propose modeling the resource usage of applications as a set of $n$ sampled values for each interference index – $I_1$, $I_2$, $I_3$ and $I_4$. Formally, the resource usage of an application $A$ is a tuple $\langle s_A, Y \rangle$, where:

- $s_a$ is a constant and called the sampling period of $A$.
- $Y$ is a $n \times 4$ matrix. We denote $y_{ij}$ as an element in $Y$. Each $y_{ij}$ represents the sampled value of the $j^{th}$ interference index at time $i \cdot s_A$.

Given the previous definition, we can define the **interference profile function** $f_{Ai}$ for each index as an interpolation between the sampled points as follows:

$$\forall i \in [1, n], \forall j \in [1, 4], f_{Aj}(i \cdot s_A) = y_{ij}$$

Without loss of generality, we consider a linear interpolation function comprised of a concatenation of linear interpolants between each pair of $(x_i, y_i)$ (Equation 17). Figure 4 depicts several examples of the profile functions for different applications.

$$f_{Aj}(x) = y_{ij} + (x - (i \cdot s_A)) \frac{y_{i+1,j} - y_{ij}}{s_A} \quad (17)$$

Given two applications $A$ and $B$, we say that they are co-scheduled if they are executed on the same physical machine; hence, they are going to share computational resources. We denote the co-scheduling operator as $\otimes$. The result of co-scheduling two applications can be interpreted as a compound application $C$. The profile function of $C$ is a combination of the profile functions of $A$ and $B$. Our key focus is to estimate the execution time of $C$ given the profiles of applications $A$ and $B$. In our model, $A$ represents an incoming application to the system, and $B$ can model the entire resource utilization of applications executed on a particular machine.

We denote by $T_A$ the execution time of application $A$; and $T_{A \otimes B}$ the execution time of application $A$ when is co-scheduled with application $B$. If application $B$ is one of the three benchmarks, we denote $T_{A \otimes B_i} = T_{AB_i}, i \in [1, 3]$. In general, $T_{A \otimes B} \neq T_{B \otimes A}$.

### B. PHASE 1: GETTING INTERFERENCE PROFILES

We consider a homogeneous computational cluster on which an application is scheduled to execute alone in a machine (this is undertaken for each applications considered in Section V-A). The interference profile functions are built using the methodology presented in Section V. An application interference profile is the set of the four interference profile functions. Figure 4 shows several examples of these application interference profiles. In addition, we obtain the total execution time when the application is executed with all available resources, denoted by $T_A$. The sample period of the profiles is denoted by $s_A$.

### C. PHASE 2: BENCHMARKING THE APPLICATION

We execute the application on a machine with different running benchmarks. In each instance we execute application $A_p$ with each application $B_j$ (e.g. povray, iozone, stream) to obtain $T_k$. Additionally, we split the execution time of application $A$ in $\lceil T_A/s_A \rceil$ intervals. The interval bounds are defined by using the number of instructions observed with perftool – #$Inst$ variable – in each time $\lceil i \cdot s_A \rceil$, $i \in [0, \lceil T_A/s_a \rceil]$ of the phase 1. We measure the time that application $A$ takes to execute the number of instructions, corresponding to the $i^{th}$ interval in phase 1, when it is co-scheduled with benchmark $B_j$. This time is denoted by $\tau_{i,j}$, $i \in [0, \lceil T_A/s_A \rceil]$, $j \in [1, 3]$. Note that $\tau_{\lceil T_A/s_A \rceil, i} = T_{AB_i}$. Interference $\delta_{i,j}$ can be computed using 18.

$$\forall i \in [1, \lceil T_A/s_A \rceil], j \in [1, 3], \quad \delta_{i,j} = \frac{\tau_{i,j} - \tau_{i-1,j}}{s_A}$$

$$\forall j \in [1, 3], \quad \delta_{0,j} = \frac{\tau_{0,j}}{s_A} \quad (18)$$

### D. PHASE 3: DEFINING THE REGRESSION MODEL

We can model the interference of an application $A$ when it is co-scheduled with an application $B$ as a multiple linear function (Equation 19)

$$\Delta(f_{A1}, \ldots, f_{A4}, f_{B1}, \ldots, f_{B4})$$
$$= y = \beta_0 + \sum_{i=1}^{4} (\beta_i f_{Ai} + \beta_{i+4} f_{Bi}) \quad (19)$$

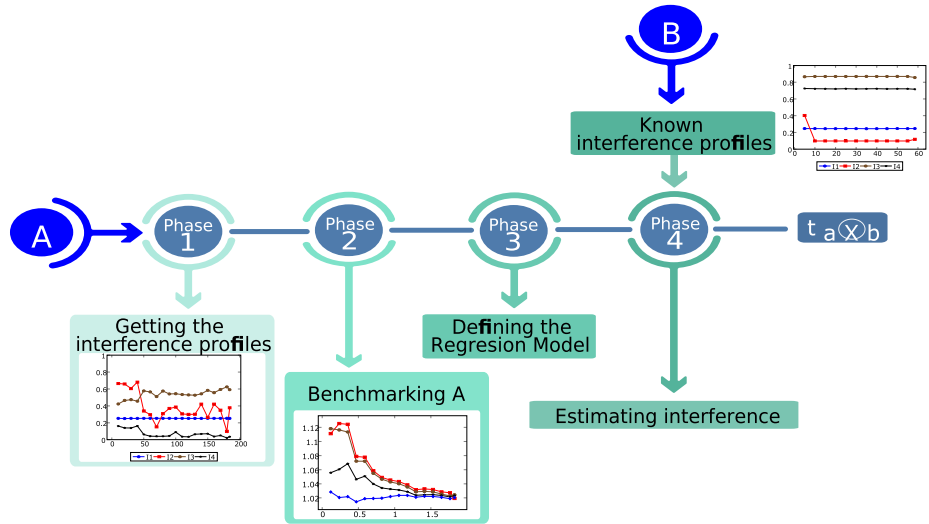

**FIGURE 5.** Methodology to estimate the interference when application *A* is co-scheduled with application *B*.

where $f_{A1}, f_{A2}, f_{A3}$ and $f_{A4}$ are values from the profile functions of *A* (Equation 17) and $f_{B1}, f_{B2}, f_{B3}$ and $f_{B4}$ are values from the profile functions of *B* at the same time.

With the interference reference values, $\delta_{ij}$ in the previous step, and the profile functions of $A$, $B_1$, $B_2$ and $B_3$, we can estimate the parameters of the model. To improve the accuracy of the model, we propose building a model for each application. To accomplish this, we must execute the application at the same time that each benchmark is executed before building the model.

As the four indices are normalized to the range [0, 1], and the interference value is positive, we can include the following restriction to the regression problem: $\beta_i > 0, \forall i \in [0, 9]$. We use the Non-Negative Least Squares (NNLS) approach [45] to estimate the regression model under these assumptions. The use of the NNLS algorithm (using the `nnls` implementation in the R package) leads to non-normal distributed residuals, so their classical interpretation should be avoided [46].

### E. PHASE 4: ESTIMATING THE INTERFERENCE

Once the interference is modeled as a linear function, we can estimate the interference of application *A* when co-scheduled with any other application *B* whose resource utilisation functions are known. The execution time of application *A* was split into $\lceil T_A/s_A \rceil$ intervals in the second phase. At instant $i \cdot s_A$ (the upper bound of interval $i$) the interference is denoted by $\delta_i$, and we use the hat notation for the estimation $\hat{\delta}_i$. Without loss of generality, the values within the interval are calculated using a linear interpolation function. Using Equation 19, we can compute the interference for each interval by using the profile function. Given a model $\Delta$ or equivalently the *beta* coefficients in the regression model, the interference for the interval $i$ can be estimated using 20.

$$\hat{\delta}_i = \Delta(f_{A1}(is_A), \ldots, f_{A4}(is_A), f_{B1}(is_A), \ldots, f_{B4}(is_A))$$
$$= \beta_0 + \beta_1 f_{A1}(is_A) + \ldots + \beta_4 f_{A4}(is_A)$$
$$+ \beta_5 f_{B1}(is_A) + \ldots + \beta_8 f_{B4}(is_A) \quad (20)$$

As the indices are functions that depend on the execution time of the application, namely their domain is $[0, T_A]$; we can rewrite Equation 20 as seen in Equation 21. To simplify this process, we compute the values of the function at the upper bound of each interval.

$$\Delta(t) = \beta_0 + \beta_1 f_{A1}(t) + \ldots + \beta_4 f_{A4}(t)$$
$$+ \beta_5 f_{B1}(t) + \ldots + \beta_8 f_{B4}(t) \quad (21)$$

The execution time of application *A* co-scheduled with application *B* can be computed as the integral of function $\Delta(t)$ (Equation 22). These can be estimated or measured values. In Equation 22, we considered that the interference is constant in each *i* interval due to sampling. Note that if we do not take any intermediary sample point $-T_A = s_A-$ we consider the interference to be constant, and the overall estimation error is likely to be higher.

$$T_{A \otimes B} = \int_0^{T_A} \Delta(t) dt \approx \sum_{i=1}^{n} \hat{\delta}_i s_A \quad (22)$$

This estimation approach can be used to develop a scheduler for a container management system. When an application arrives, the scheduler calculates the interference functions and selects the best machine to deploy the application. When new applications arrive, the scheduler knows the profile of the application scheduled on each machine and can estimate the machine that leads to the lowest interference and, consequently, to the shortest execution time.

**TABLE 1.** Overall interference for applications when they are co-scheduled with benchmarks.

| App. | $T_A(s)$ | $T_{AB_1}(s)$ | $T_{AB_1}/T_A$ | $T_{AB_2}(s)$ | $T_{AB_2}/T_A$ | $T_{AB_3}(s)$ | $T_{AB_3}/T_A$ |
|---|---|---|---|---|---|---|---|
| metis | 86.01 | 114.37 | 1.33 | 114.22 | 1.33 | 193.67 | 2.25 |
| montage | 374.08 | 397.63 | 1.06 | 421.73 | 1.13 | 401.41 | 1.07 |
| bzip2 | 64.52 | 69.05 | 1.07 | 66.77 | 1.03 | 78.75 | 1.22 |
| pbzip2 | 20.0 | 37.94 | 1.9 | 20.26 | 1.01 | 36.16 | 1.81 |
| blastn | 155.25 | 171.26 | 1.1 | 157.98 | 1.02 | 170.71 | 1.1 |
| blastx | 180.0 | 190.02 | 1.06 | 184.33 | 1.02 | 217.55 | 1.21 |

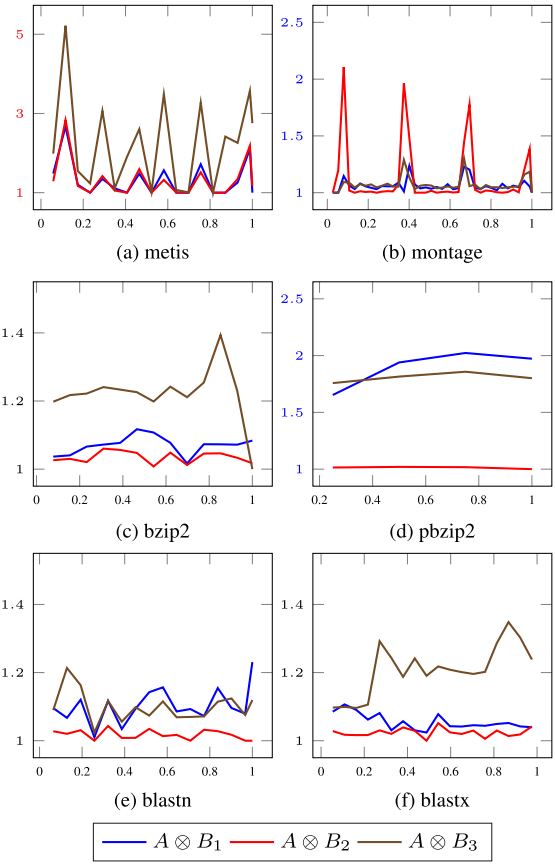

**FIGURE 6.** Interference values vs. normalize application execution time (*A*) –metis (a), montage (b), bzip2 (c), pbzip2 (d), blastn (e) and blastx (f) –when they are co-scheduled with the benchmarks – $B_1$ is pov-ray, $B_2$ is iozone and $B_3$ is stream.

## VII. EXPERIMENTAL EVALUATION

All experiments were executed on eight homogeneous physical machines with the same hardware and software configuration. On the hardware side, each machine has an Intel i5-4690 (3,500GHz) CPU with four cores, 32 GB of RAM and 1 TB hard disk at 7200rpm. On the software side, the configuration was minimal to not interfere with our experiments. Each machine has a minimal Ubuntu 16.04 server operating system with only basic services (ssh and docker services). Docker (version 18.02.0-ce) ran the applications of the experiments in containers, and no other application ran on the machine.

Thus, negligible interference was expected on the platform used for all experiments.

The sampling period ($s_A$) was 5 seconds. Each experiment was run seven times to get a 95% confidence interval that is within 5% of the mean. In total, 515 samples were obtained from each execution. Each sample had 11 values corresponding to the 11 basic variables described in Subsection V-A.

All experiments presented in this paper were run with this setup.

In this section, we present additional experiments that were conducted to test the proposed methodology. First, we present the results of the benchmarking phase and interference estimation.

### A. BENCHMARKING PHASE

Figure 6 depicts the interference of different applications when they are co-scheduled with the three benchmark applications. The x-axis shows the execution time of the application normalized to one and the y-axis represents the interference as computed in Equation 18. The y-axis scale for bzip2, blastn and blastx is different to better visualization the impact of interference in each of these instances. We can see that metis has a similar behavior to bzip2 because both of them are applications which make a high usage of memory bandwidth. In both cases, the highest interference is achieved when they are co-scheduled with stream, which makes an aggressive usage of memory bandwidth. Although, metis can produce some phases, at the beginning and in the end, with higher CPU and cache misses.

On the other hand, pbzip2 is highly affected by pov-ray, because both applications use a CPU fully, and by stream, because of its high memory bandwidth. The Montage application ()Figure 6b) shows several well-differentiated phases,with regular high resource usage, mainly cache misses observed affected by iozone. Each of these phases corresponds to the task of the montage workflow. The Blastn and blastx applications are tolerant/resilient to the benchmark application interference, with low degradation in both cases.

Table 1 shows the overall degradation for the six applications. It can be interpreted as the mean value of the interference shown in Figure 6. As expected, parallel applications –such as pbzip2– suffer a higher degradation when they are co-scheduled with $B_1$ than one-core applications. The interference caused on the latter by this benchmark is about 5%-7%.

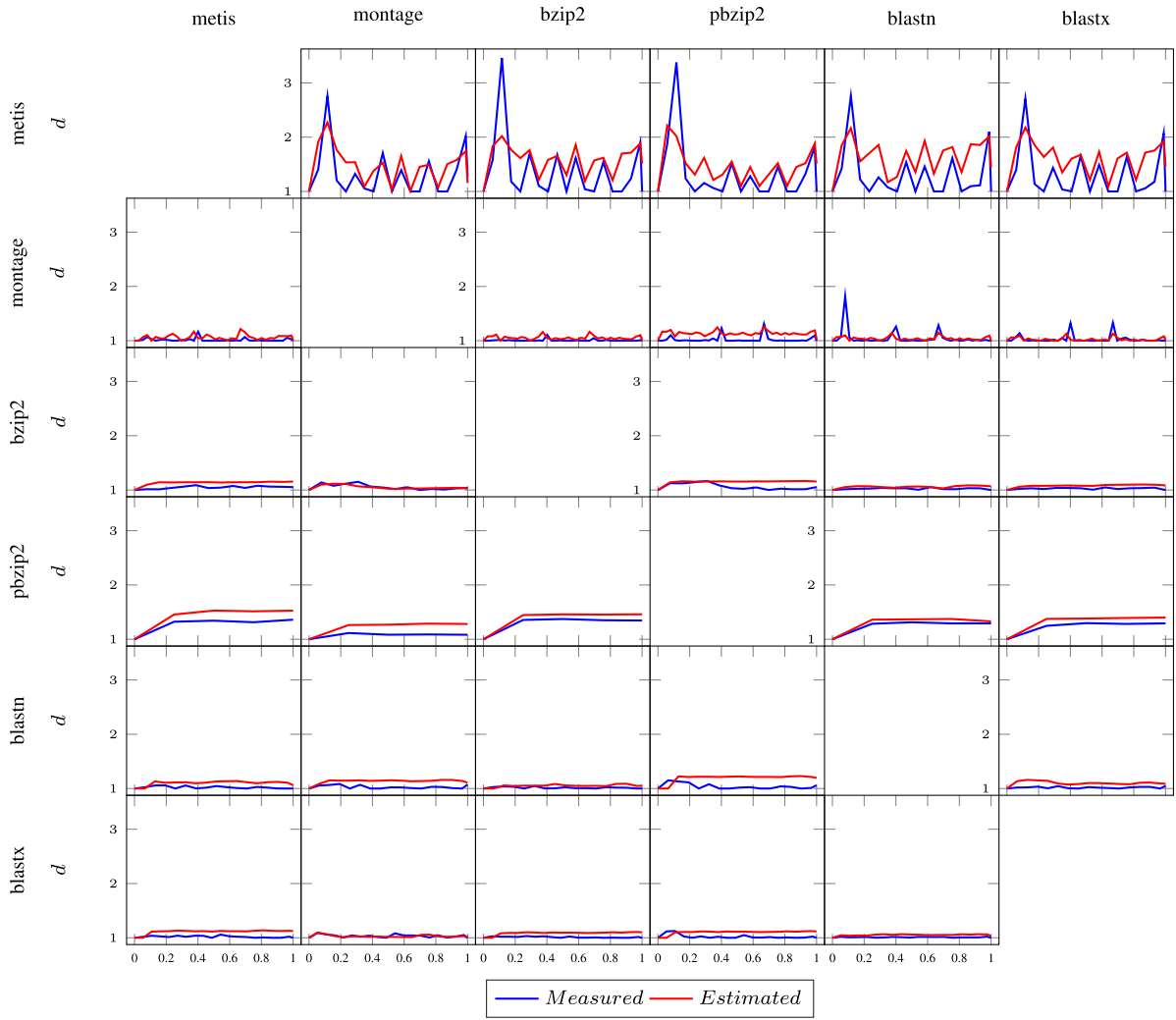

**FIGURE 7.** Interference values vs. normalized execution time for several co-scheduled applications ($A \otimes B$). Blue line shows the measured values and Red line shows the estimated values with the proposed methodology.

### B. EXPLOITING THE MODEL

We measured the interference observed when the previously chosen applications were co-scheduled with others. The results are shown in figure 7. The red line, in the results, represents the value estimated using the proposed methodology.

This figure shows that our prediction model can provide meaningful estimations (in blue), near the measured temporal behavior (in red), for applications with distinctive resource usage. The biggest interference, which can be observe, corresponds to when metis is executed alongside the rest of the applications, because of the important usage of resources, mainly memory bandwidth, as shown in figure 6. The estimation of some spikes in the measured interference, could be calculated if a shorter sampling period was used in the benchmarking phase, as we analyze in the next section. This same consideration can be made to some interference spikes between montage and blastn applications, and some other combinations and periods, as shown in figure 7. Most often, the estimations observed in this figure, offers lightly larger interference values, as a cautious approach. Overall, the estimated and measured interferences followed a similar pattern, validating our approach.

Table 2 depicts the Mean Error (*ME*), the Mean Squared Error (*MSE*) and the accuracy of estimations (*Acc*). The first ones have been calculated using Equations 23 and 24. Although the regression methodology tries to minimize the absolute error, Mean Relative Error (*MRE*) provides a more useful metric to analyze the accuracy of the model (Equation 25).

Once we have estimated the interference for each interval, we compute the execution time (Eq. 22). Table 3 lists the values. The estimated execution time is computed as the sum of all the estimated interference, and the residual error $\epsilon$ in the estimation depends on the *ME* and the sampling parameters, namely, the number of sampling points $n$ and sampling period $s_A$. The expression is given by Eq. 26.

TABLE 2. Mean Error –ME–, Mean Squared Error –MSE– and Mean Accuracy –Acc– values for the experiment of figure 7.

|         |     | metis | montage | bzip2 | pbzip2 | blastn | blastx | Total |
|---------|-----|-------|---------|-------|--------|--------|--------|-------|
| **metis**   | $ME$  | -     | -0.14   | -0.21 | -0.14  | -0.34  | -0.29  | -0.22 |
|         | $MSE$ | -     | 0.1     | 0.25  | 0.16   | 0.22   | 0.18   | 0.18  |
|         | $Acc$ | -     | 0.8     | 0.71  | 0.79   | 0.65   | 0.69   | 0.73  |
| **montage** | $ME$  | -0.05 | -       | -0.05 | -0.11  | 0.0    | -0.0   | -0.04 |
|         | $MSE$ | 0.01  | -       | 0.01  | 0.02   | 0.02   | 0.01   | 0.01  |
|         | $Acc$ | 0.94  | -       | 0.94  | 0.88   | 0.95   | 0.97   | 0.94  |
| **bzip2**   | $ME$  | -0.08 | 0.0     | -     | -0.09  | -0.03  | -0.05  | -0.05 |
|         | $MSE$ | 0.01  | 0.0     | -     | 0.01   | 0.0    | 0.0    | 0.0   |
|         | $Acc$ | 0.92  | 0.98    | -     | 0.92   | 0.97   | 0.95   | 0.95  |
| **pbzip2**  | $ME$  | -0.14 | -0.15   | -0.08 | -      | -0.05  | -0.09  | -0.1  |
|         | $MSE$ | 0.02  | 0.03    | 0.01  | -      | 0.0    | 0.01   | 0.01  |
|         | $Acc$ | 0.9   | 0.87    | 0.94  | -      | 0.96   | 0.93   | 0.92  |
| **blastn**  | $ME$  | -0.04 | -0.04   | -0.05 | -0.04  | -      | -0.05  | -0.04 |
|         | $MSE$ | 0.0   | 0.0     | 0.0   | 0.01   | -      | 0.0    | 0.0   |
|         | $Acc$ | 0.95  | 0.96    | 0.95  | 0.94   | -      | 0.95   | 0.95  |
| **blastx**  | $ME$  | -0.09 | 0.0     | -0.07 | -0.08  | -0.04  | -      | -0.06 |
|         | $MSE$ | 0.01  | 0.0     | 0.01  | 0.01   | 0.0    | -      | 0.01  |
|         | $Acc$ | 0.91  | 0.98    | 0.92  | 0.91   | 0.96   | -      | 0.94  |
| **Total**   | $ME$  | -0.08 | -0.07   | -0.09 | -0.09  | -0.09  | -0.1   | -0.09 |
|         | $MSE$ | 0.01  | 0.03    | 0.06  | 0.04   | 0.05   | 0.04   | 0.04  |
|         | $Acc$ | 0.92  | 0.92    | 0.89  | 0.89   | 0.9    | 0.9    | 0.9   |

TABLE 3. Measured and Estimated execution times in seconds –Mea. and Est. rows– for the experiment of Figure 7.

|         |      | metis  | montage | bzip2  | pbzip2 | blastn | blastx |
|---------|------|--------|---------|--------|--------|--------|--------|
| **metis**   | Mea. | -      | 115.36  | 121.13 | 116.49 | 114.30 | 114.96 |
|         | Est. | -      | 128.94  | 138.95 | 127.46 | 144.44 | 140.84 |
| **montage** | Mea. | 379.72 | -       | 376.66 | 383.55 | 392.83 | 385.16 |
|         | Est. | 399.71 | -       | 396.07 | 427.39 | 391.83 | 386.90 |
| **bzip2**   | Mea. | 67.79  | 68.27   | -      | 68.73  | 66.08  | 66.31  |
|         | Est. | 73.59  | 67.94   | -      | 74.65  | 68.48  | 69.97  |
| **pbzip2**  | Mea. | 26.70  | 21.85   | 27.12  | -      | 25.91  | 25.60  |
|         | Est. | 30.13  | 25.49   | 29.10  | -      | 27.15  | 27.74  |
| **blastn**  | Mea. | 158.62 | 159.55  | 157.62 | 161.54 | -      | 158.04 |
|         | Est. | 165.71 | 166.35  | 165.72 | 167.62 | -      | 166.45 |
| **blastx**  | Mea. | 188.16 | 190.94  | 186.69 | 188.60 | 186.33 | -      |
|         | Est. | 206.02 | 190.19  | 200.94 | 203.92 | 194.44 | -      |

$$ME = \frac{1}{n}\sum_{i=1}^{n}(\delta_i - \hat{\delta}_i) \qquad (23)$$

$$MSE = \frac{1}{n}\sum_{i=1}^{n}(\delta_i - \hat{\delta}_i)^2 \qquad (24)$$

$$Acc = 1 - MRE = 1 - \frac{1}{n}\sum_{i=1}^{n}\frac{|\delta_i - \hat{\delta}_i|}{\delta_i} \qquad (25)$$

$$\epsilon = T_A - \hat{T}_A = \sum_{i=1}^{n}\delta_i s_A - \sum_{i=1}^{n}\hat{\delta}_i s_A$$
$$= \sum_{i=1}^{n}(\delta_i s_A - \hat{\delta}_i s_A) = n s_A \sum_{i=1}^{n}(\delta_i - \hat{\delta}_i)$$
$$= n^2 \cdot s_A \cdot ME \qquad (26)$$

## VIII. REFINING THE MODEL

In this section, we analyze how the methodology can be refined by considering the experimental results. First, we evaluated the impact of sampling time on the accuracy of the estimated execution time. Second, we removed the benchmarking phase and evaluated a single model for all the arriving applications. Finally, we discuss how our methodology can be exploited by the scheduler of the container management system.

**TABLE 4.** Measured and Estimated execution times in seconds when Metis is co-scheduled with Montage. Two estimated scenarios. In Scenario 1, the sampling period of Phases 1 and 3 is always 5 seconds, and Phase 4 applies the different sampling periods defined in the $s_A$ column in seconds. In Scenario 2, $s_A$ is the same for all phases and corresponds to the values in its column. $n$ is the number of sampling intervals.

| $n$ | $s_A$ | Measured | Scenario 1 | Scenario 2 |
|---|---|---|---|---|
| 17 | 5 | 115.36 | 128.94 | 128.94 |
| 9 | 10 | 115.36 | 132.182 | 154.53 |
| 5 | 20 | 115.36 | 133.86 | 158.52 |
| 3 | 40 | 115.36 | 143.9 | 154.6 |
| 1 | 85 | 115.36 | 159.94 | 135.34 |

### A. SAMPLING TIME

In our model, we consider that performance losses are not constant during the execution time of an application. The accuracy of these values depends on the sampling time and interpolation function. In the previous sections, we considered a sampling period of 5 seconds for all applications.

However, if we consider the experiment with Metis and Montage, $metis \otimes Montage$, we can compute the accuracy of the estimated execution time for a given sample period. In table 4, we can observe the results of the two scenarios of execution of the experiment. On the one hand, Scenario 1 computed all the profiles (phases 1 and 3) with a 5 seconds of sample period ($s_A$), but the overall interference (phase 4) was executed with the different values of the sample period defined in column $s_A$. On the other hand, Scenario 2 applies the values in column $s_A$ to all phases, thus, each sample value is applied, either for the profiling phases or the overall interference phase.

For simplicity, we assumed that the sampling period is constant in all phases of the methodology; however, we can use one sampling period to compute the profiles and an another one to compute the overall interference (Scenario 1 in Table 4) and the execution time (Scenario 2 in Table 4).

We observe that the accuracy of the estimation increases as the number of sampling period increases in Scenario 1. The reason for this behavior is straightforward. As we increase the number of intervals, the size of those intervals decreases, thus, the error estimation also decreases. For Scenario 2, the behavior was quite similar. We reduced the number of points to interpolate the interference profile of the applications, consequently, the estimation was less fine-grained.

The experimental results lead us to conclude that the sampling period should be adjusted depending on the granularity needed for the estimations and the expected accuracy. Namely, if we have a scheduler that tries to fill the gaps of low expected interference values in the execution of long applications with smaller applications, the granularity can determine the size of these gaps and the sampling period.

### B. SINGLE MODEL

In our methodology, we built a model for each incoming application to make estimations. This model allows us to wait for the execution time of the application when it is co-scheduled with the three benchmarks. This is easily parallelisable, and the waiting time of the benchmarking phase, $T_{bench}$ is given by Equation 27.

$$T_{bench} = max_{i=1}^{i=3}\{T_{B_i}\} \qquad (27)$$

This time can be negligible if we consider scenarios in which many similar applications arrive at the system to be processed, as occurs in a lambda function processing architecture. However, we can analyze what happens when we build a single model using the values of all experiments and when we use that model to estimate the execution time.

For example, we can consider a single model that includes the interference values of metis, bzip2, pbzip2 and blastn applications when they are co-scheduled with the benchmarking applications. We excluded montage and blastx applications to analyze the accuracy of the model when an application that did not include arrives. Results are shown in Table 5. Note that the accuracy value –Acc– is the mean value of the accuracy of all estimations for that experiment (see Equation 25) and, not the accuracy of the total estimated time. The overall accuracy was approximately 0.78. This value includes the estimation of co-scheduling all applications with montage and blastx, and the $montage \otimes blastx$ experiment in which both applications are not in the model.

These results show that the proposed single model captures the variability of the variables quite well and can be used as an upper bound of the real value. Additionally, it seems that applications that are not included in the model do not exhibit worse behavior than those that are included. These results lead us to conclude that the benchmarking phase improves the overall accuracy of the model at the expense of waiting for the benchmarking time. Thus, the single model can be used if an estimation of the upper bound is required and/or if its accuracy is high enough.

### C. EXPLOITING THE MODEL

In the previous sections, we focused on estimating the effects of the interference between containers. We have defined the co-scheduling operator $\otimes$ and we have calculated the execution time. These values are useful for the scheduler to determine the optimal host for deploying the container. Several scheduling techniques (e.g. the use of priority queues) can be used to improve the performance of executed applications. However, with our methodology, the scheduler can also determine when the best time to launch the application is.

We can define the co-scheduling operator with a delay $\otimes^k$. $A \otimes^k B$ denotes that application $A$ is co-scheduled with application $B$; however, $A$ is delayed until $B$ reaches the $k$ interval. This approach can be useful for executing short applications when long execution applications are in a low resource-usage period. The scheduler can choose the machine

TABLE 5. Measured and Estimated execution times in seconds –Mea. and Est. rows– and Accuracy –*Acc*– with a single model which includes metis, bzip2, pbzip2, blastn applications.

|         |      | metis  | montage | bzip2  | pbzip2 | blastn | blastx | Total |
|---------|------|--------|---------|--------|--------|--------|--------|-------|
| **metis**   | Mea. | -      | 115.36  | 121.13 | 116.49 | 114.3  | 114.96 | -     |
|         | Est. | -      | 117.92  | 121.0  | 144.3  | 125.21 | 122.63 | -     |
|         | *Acc* | -     | 0.8     | 0.74   | 0.59   | 0.73   | 0.74   | 0.72  |
| **montage** | Mea. | 379.72 | -       | 376.66 | 383.55 | 392.83 | 385.16 | -     |
|         | Est. | 480.4  | -       | 427.67 | 532.15 | 430.97 | 436.09 | -     |
|         | *Acc* | 0.74  | -       | 0.87   | 0.62   | 0.87   | 0.85   | 0.79  |
| **bzip2**   | Mea. | 67.79  | 68.27   | -      | 68.73  | 66.08  | 66.    | -     |
|         | Est. | 87.74  | 76.34   | -      | 95.18  | 79.39  | 79.48  | -     |
|         | *Acc* | 0.73  | 0.89    | -      | 0.64   | 0.81   | 0.82   | 0.78  |
| **pbizp2**  | Mea. | 26.7   | 21.85   | 27.12  | -      | 25.91  | 25.6   | -     |
|         | Est. | 34.18  | 32.47   | 31.24  | -      | 31.94  | 32.3   | -     |
|         | *Acc* | 0.78  | 0.61    | 0.88   | -      | 0.81   | 0.79   | 0.77  |
| **blastn**  | Mea. | 158.62 | 159.55  | 157.62 | 161.53 | -      | 158.04 | -     |
|         | Est. | 200.16 | 177.91  | 179.18 | 219.39 | -      | 184.61 | -     |
|         | *Acc* | 0.75  | 0.88    | 0.86   | 0.64   | -      | 0.84   | 0.79  |
| **blastx**  | Mea. | 188.16 | 190.95  | 186.69 | 188.6  | 186.34 | -      | -     |
|         | Est. | 226.28 | 195.45  | 199.71 | 249.01 | 212.26 | -      | -     |
|         | *Acc* | 0.81  | 0.95    | 0.93   | 0.68   | 0.87   | -      | 0.85  |
| **Total**   | *Acc* | 0.76  | 0.83    | 0.86   | 0.63   | 0.82   | 0.81   | 0.78  |

and delay that are optimal to minimize the interference caused by the applications in the cluster.

In addition, interference profiles can be useful for analyzing the behavior of applications. For example, they can be used to give penalties, or rewards, to those applications that interfere more, or less, with the remaining containers. Although several resources are difficult to isolate, our model can analyze the impact of setting up limits for resources that can be isolated (e.g., CPU, I/O disk, or network usage). In this regard, several container management systems such as Kubernetes allows bounds to be set for the CPU used by a container.

## IX. CONCLUSION

Co-located container hosted applications can suffer from performance interference, particularly if such applications have similar resource requirements over their execution lifetime. This interference is caused by contention across different low-level resources, such as CPU, cache and memory hierarchy (leading to page faults), which can be difficult to characterize fully. We proposed a performance interference model to predict the performance degradation of co-scheduled applications hosted in containers. The model considers that an application makes use of the CPU, memory, and I/O and these requirements can vary over the execution lifetime of the container (and the application it hosts). Using Confirmatory Factor Analysis (CFA), we identified a set of high-level, human-comprehensible analytical indices that characterize how a service uses the physical resources of a machine. These indices are expressed as time series, so that we can analyze how they interfere across the execution lifetime of an application, instead of considering them as a constant parameter that is only measured once. We subsequently use a multiple linear regression model to estimate how much a service is going to be degraded for co-scheduling with others. Our experiments were conducted using Docker containers with a number of real applications of varying needs. Experiments suggest that for different service combinations, our model provides an improved representation of their performance variability to make accurate estimations of overall execution time. Thus, the proposed interference model can be used to decrease executions costs of applications and improve their performance and overall resources usage in dynamic Cloud infrastructures. Moreover, these indices can be used with a low overhead in future scheduling proposals.

The proposed approach can also be used as the basis for developing interference-tolerant applications. An application could dynamically adapt its behavior if it noticed meaningful interference from other co-hosted applications. Interference-awareness of this kind can subsequently be used to guarantee minimal performance targets for applications.

In future work, the use of the four interference indices proposed in this study will be explored as a criterion to improve the scheduling of distributed jobs, services, or applications across resources within a cloud environment. Moreover, additional resources will be considered, such as GPU, network usage and I/O file system access.

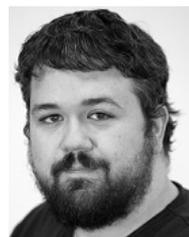

**VÍCTOR MEDEL** received the B.S., M.S., and Ph.D. degrees in computer science from the University of Zaragoza, Spain. He is the Head of Digital Platforms in Nervia Consultores, an IT service and consulting company. His interests include modeling and analyzing the performance of distributed systems and DevOps for deploying in distributed systems.

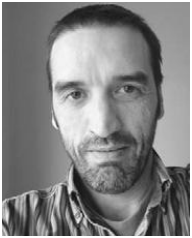

**UNAI ARRONATEGUI** received the B.S. degree in electronic engineering from the Escuela Politécnica de Mondragón, Arrasate, Spain, and the M.S. and Ph.D. degrees in computer science from Université Paul Sabatier, Toulouse, France.

He is currently an Associate Professor with the Universidad de Zaragoza, Zaragoza, Spain. His research interests include distributed systems, computer networks, and system administration.

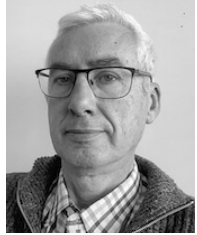

**JOSÉ ÁNGEL BAÑARES** was born in Zaragoza, Spain, in 1966. He received the M.S. degree in industrial electrical engineering and the Ph.D. degree in computer science from the University of Zaragoza, in 1991 and 1996, respectively. He joined the Faculty of the University of Zaragoza, as an Assistant Professor, in 1994, where he has been an Associate Professor, since 1999. His research interests include petri nets, artificial intelligence, and distributed computing.

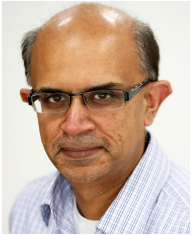

**OMER RANA** received the Ph.D. degree in neural computing and parallel architecture from the Imperial College, University of London. He is currently a Professor of performance engineering with the School of Computer Science and Informatics, Cardiff University. His research interests extend to three main areas within computer science: high performance distributed systems, data analysis, and multi-agent systems. He has also used these approaches in a number of application areas.

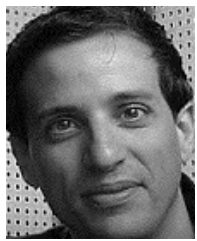

**RAFAEL TOLOSANA-CALASANZ** is currently an Associate professor with the Computer Science and Systems Engineering Department, University of Zaragoza. His research interests include intersection of distributed and parallel systems and problem solving environments.

• • •